\documentclass[reprint,floatfix,superscriptaddress,amsmath,amssymb,aps,showkeys]{revtex4-2}
\usepackage{xcolor}
\usepackage{graphicx}
\usepackage{subcaption}
\usepackage{dcolumn}
\usepackage{bm}
\usepackage{braket}
\usepackage{amsthm}
\usepackage{hyperref}
\usepackage{hyperref}
\hypersetup{
	colorlinks=true,
	linkcolor=blue,
	citecolor=blue,
	urlcolor=blue}
\usepackage{float}
\usepackage{graphicx}
\usepackage{subcaption} 

\begin{document}

\sloppy

\title{Exact decoherence dynamics of a detector coupled to \\vacuum fluctuations in bounded geometries}

\author{Zi'an Zhao}

\affiliation{Institute of Fundamental Physics and Quantum Technology, Zhejiang Key Laboratory of Extreme Universe, \\
and School of Physical Science and Technology, 
Ningbo University, Ningbo, Zhejiang 315211, China}
\affiliation{Department of Physics, Fudan University, Shanghai, 200433, China}

\author{Wenting Zhou}
\email{zhouwenting@hunnu.edu.cn}
\affiliation{Institute of Fundamental Physics and Quantum Technology, Zhejiang Key Laboratory of Extreme Universe, \\
and School of Physical Science and Technology, Ningbo University, Ningbo, Zhejiang 315211, China}

\date{\today}

\begin{abstract}
The properties of quantum vacuum are highly sensitive to the underlying spacetime geometry and boundary conditions. We investigate in this paper the decoherence dynamics of a two-level system (detector) coupled to a vacuum scalar field through the $\hat{\sigma}_z$-coupling scheme under various boundary environments. By rigorously evaluating the reduced density matrix of the detector, we derive an exact general expression for the detector's decoherence function, $\gamma(\Delta t)$, and use it to characterize the dynamical decoherence of a detector in three scenarios: free space, a perfectly reflecting boundary, and a cavity formed by two parallel reflecting boundaries. We find that in free space the detector's coherence decays monotonically with time, indicating that vacuum fluctuations alone inevitably lead to a complete loss of coherence. In the single-boundary case, by contrast, the detector's coherence displays non-monotonic time evolution: $\gamma(\Delta t)$ increases with time up to a maximum and then settles to a steady value, indicating that the initial coherence can be partially preserved due to the presence of the boundary. When the detector is placed inside a cavity, its coherence evolution exhibits oscillatory behavior, and a significant fraction of the initial coherence survives. Moreover, for the same detector-boundary distance, more coherence is preserved in the cavity case than in the single-boundary case. These results provide a  viable route for engineering quantum coherence via boundary conditions.   

\end{abstract}

\maketitle


\section{Introduction}\label{sec:intro}

The properties of quantum vacuum lie at the heart of quantum field theory (QFT). Unlike the classical notion of empty vacuum, the QFT vacuum is teeming with fluctuations. Owing to its ubiquitous coupling to quantum fields, any physical system can never be perfectly isolated but is continuously perturbed by these vacuum fluctuations. As a result, when a quantum system is initially prepared in a superposition state, its phase coherence is inevitably lost. This phenomenon, known as the environment-induced decoherence (decoherence for short), is widely  recognized as the origin of the quantum-to-classical transition~\cite{Joos85,Zurek86,Zurek91}. During the past decades, decoherence has been extensively investigated, as it plays a crucial role in fields ranging from atomic physics, quantum information processing to surface science, quantum gravity and quantum biology 
(for reviews, see Refs.~\cite{Paz01,Zurek03,Joos03,Schlosshauer05,Schllosshauer07,Fischer09,Paladino14,Moran19}). 

A fundamental property of the QFT vacuum is that it is highly sensitive to the underlying spacetime geometry. In this regard, the coupling to quantum fields constitutes an intrinsic and universal source of decoherence that cannot be eliminated by conventional shielding. The model of a localized two-level system (often referred to as an  Unruh-DeWitt detector) interacting with a quantum field serves as a simple yet powerful theoretical framework to probe these vacuum properties. For instance, this model has been instrumental in revealing fundamental phenomena such as the Fulling-Davies-Unruh (FDU) effect, where the acceleration modifies the detector's decoherence rate due to the apparent thermal property of Rindler vacuum~\cite{Nesterov20,Xu23,Liu25,Sahota26}.
These environment-dependent decoherence behaviors, relative to an inertial detector in an unbounded vacuum, originate from the reshaping of the field modes. 

Boundaries are another key factor capable of reshaping the field modes, thereby resulting in a variety of observable quantum phenomena, such as the Casimir effect~\cite{Casimir48,Lifshitz55,Milonni94,Bordag01}, the modified dynamical evolution and radiative properties of quantum emitters near conducting plates, dielectric interfaces, or cavities~\cite{Ochiai06,Spagnolo06,Xu09,Manassah11,Wu15,Arias16,Amooghorban17,Zhou18,Chen26} (just to name a few). 
It is therefore natural to expect alterations in the decoherence properties of a quantum system located in a bounded environment. Indeed, such investigations have already been carried out. By calculating the $l_1$ norm and relative entropy of coherence for a static two-level quantum system dipole-coupled to a thermal environment, it was found that the decoherence near planar boundaries, compared with the free-space case, can be effectively inhibited by tuning the system's polarizability or the system-boundary separation~\cite{Liu16}. Moreover, the interplay between thermal noise, planar or cylindrical boundaries, and acceleration in determining the 
decoherence of a two-level system coupled to a scalar field has also been investigated. It was found that the FDU effect in a suitably engineered bounded environment can enhance rather than degrade quantum coherence~\cite{Liu25,Sahota26}. Recently, studies of boundary effects on decoherence have also been extended from the single-qubit case to the multi-qubit case~\cite{Wu26}.  

However, almost all existing studies of boundary effects on the decoherence of two-level quantum systems focus on a particular coupling, $\hat{H}_I\sim\hat{\sigma}_x\hat{\phi}(x)$, where $\hat{\sigma}_x$ and $\hat{\phi}(x)$ denote the Pauli-$x$ matrix and a scalar field, respectively. In this work, we are interested in the role of boundaries in the dynamical decoherence of a two-level quantum system coupled to a vacuum scalar field through a different interaction, namely, the $\hat{H}_I\sim\hat{\sigma}_z\hat{\phi}(x)$ coupling. This Hamiltonian is equivalent to the one introduced in the context of the tunneling problem~\cite{Leggett87,Gardiner91}, and it has been used to model quantum decoherence in quantum computers~\cite{Unruh95,Palma96} and in studies of the FDU effect~\cite{Nesterov20}. For brevity, we refer to these two couplings as the $\hat{\sigma}_x$-coupling and $\hat{\sigma}_z$-coupling schemes. The decoherence associated with the $\hat{\sigma}_x$-coupling is generally analyzed within the framework of open quantum system theory, often under the Born-Markov approximation~\cite{Breuer02,Rivas12}, or by means of a Dyson series expansion for the composite system. The Dyson series generally cannot be summed exactly, so a truncation at leading order in the coupling constant is often required. In contrast, the $\hat{\sigma}_z$-coupling scheme allows for an exact description of the system's evolution, without invoking the Born--Markov approximation or perturbative treatment.

In this paper, we address boundary effects on the decoherence dynamics of a two-level system (detector) coupled to a vacuum quantum scalar field via the $\hat{\sigma}_z$-coupling, considering three cases: (i) free space, (ii) a single perfectly reflecting planar boundary, and (iii) a cavity formed by two parallel perfectly reflecting boundaries. For each case, we derive exact expressions for the decoherence function, analyze the resulting dynamics, and demonstrate the possibility of preserving coherence. The paper is organized as follows. In Section II, we introduce the model of a detector linearly coupled to a quantum scalar field. In Section III, we investigate the dynamical decoherence of the system in the above three cases. We summarize our results in Section IV.  

\section{Model}\label{sec:decoherence}

Owing to the quantization of the field, vacuum fluctuations persist even in the absence of real particles. Consequently, a quantum system is never perfectly isolated and undergoes dissipative  evolution, even in vacuum. In this section, we introduce a quantitative framework for describing the dynamical decoherence of a two-level system (detector) coupled to vacuum fluctuations via the $\hat{\sigma}_z$-coupling.

We designate $|g\rangle$ and $|e\rangle$ as the two eigenstates of the detector, separated by an energy gap $\omega_0$, and describe the vacuum fluctuations in terms of a massless scalar field
\begin{equation}
\hat{\phi}(t,\mathbf{x})=\int d^3\mathbf{k}\left[\hat{a}_{\mathbf{k}}e^{-i\omega_{\mathbf{k}}t}u_{\mathbf{k}}(\mathbf{x})+ H.c.\right].
	\label{eq:averaged_field}
\end{equation}
Here $\hat{a}_{\mathbf{k}}$ is the annihilation operator of the field with wave vector $\mathbf{k}$,  $\omega_{\mathbf{k}}=|\mathbf{k}|$, and $u_{\mathbf{k}}(\mathbf{x})$ represents a complete set of normalized field modes. The vacuum state is defined as $\hat{a}_{\mathbf{k}}|0\rangle = 0$ for all $\mathbf{k}$.
%

The free Hamiltonian of the composite ``detector+vacuum" system can be expressed as
\begin{equation}
\hat{H}_0=\frac{\omega_0}{2}\,\hat{\sigma}_z \otimes \mathbb{I} + \mathbb{I} \otimes \sum_{\mathbf{k}}\omega_{\mathbf{k}}\,\hat{a}_{\mathbf{k}}^\dagger \hat{a}_{\mathbf{k}}\;,
\label{eq:free_H}
\end{equation}
where $\mathbb{I}$ is the identity operator, $\hat{\sigma}_z=|e\rangle\langle e|- |g\rangle\langle g|$ is the Pauli operator of the detector in the $z$ direction.

The detector-field interaction Hamiltonian takes the form~\cite{Unruh95}
\begin{equation}
\hat{H}_1=\lambda\,\hat{\sigma}_z\otimes\int d^3\mathbf{k}\left[\hat{a}_{\mathbf{k}} 
g_{\mathbf{k}}+ H.c.\right]\;,
\label{eq:interaction_H}
\end{equation}
where $\lambda$ is the coupling constant,  
\begin{equation}
g_{\mathbf{k}}= \int d^3\mathbf{x}\,f(\mathbf{x})u_{\mathbf{k}}(\mathbf{x})
\label{eq:g-function}
\end{equation}
with $u_{\mathbf{k}}(\mathbf{x})$ representing a complete set of field modes and
\begin{equation}
f(\mathbf{x})=\frac{l}{\pi^2(|\mathbf{x}|^2+l^2)^2}
\label{eq:form_factor}
\end{equation}
being a profile function that characterizes the spatial extent of the detector. The associated length scale $l$ is taken to be of the order of the Bohr radius. This profile function is introduced to avoid ultraviolet divergence in the decoherence rate that arises in the point-like detector limit $l\rightarrow0$. As we shall see, this yields an exponential factor $e^{-\omega_{\mathbf{k}} l}$ in the detector-field coupling, thereby rendering the decoherence function finite. Such a regularization scheme has also been adopted in studies of the response rate of an Unruh-DeWitt detector~\cite{Louko06} and of the relationship between noninertial detector decoherence and the FDU effect~\cite{Schlicht04,Nesterov20}. Alternatively, regularization can be achieved through other schemes, for instance, by introducing a frequency cut-off for the field~\cite{Palma96} or a switching function governing the duration of the detector-field interaction~\cite{Svaiter92,Higuchi93,Sriramkumar96,Satz07}.

Suppose that the detector is initially prepared in an arbitrary superposition of its ground and excited states $|\psi(t_0)\rangle = c_1|g\rangle + c_2|e\rangle$, where $c_1$ and $c_2$ are complex coefficients satisfying the normalization condition $|c_1|^2+|c_2|^2=1$. Then the state of the composite detector-field system is initially given by
\begin{equation}
|\Psi(t_0)\rangle=\left(c_1|g\rangle + c_2|e\rangle\right)\otimes |0\rangle\;,
\label{eq:initial_product_state}
\end{equation}
and its unitary time evolution in the interaction picture is governed by
\begin{eqnarray}
|\Psi(t)\rangle
=\mathcal{\hat{T}}e^{-i\int_{t_0}^{t}dt'\,\hat{H}_I(t')}|\Psi(t_0)\rangle\;,
\label{eq:rho_t}
\end{eqnarray}
where $\mathcal{\hat{T}}$ is the time ordering operator and
\begin{eqnarray}
\hat{H}_I(t)=e^{i\hat{H}_0t}\hat{H}_1e^{-i\hat{H}_0t}
=\lambda\,\hat{\sigma}_z\otimes\hat{\bar{\phi}}(t)
\label{eq:H_I}
\end{eqnarray}
with the averaged field operator defined as
\begin{eqnarray}
\hat{\bar{\phi}}(t)
&\equiv&\int d^3\mathbf{k}\left[\hat{a}_{\mathbf{k}}e^{-i\omega_{\mathbf{k}}t}g_{\mathbf{k}}+ H.c.\right]\;.
\label{eq:averaged_field}
\end{eqnarray}

To investigate the decoherence dynamics of the detector, we next examine the reduced density operator $\hat{\rho}_d(t)$, obtained by tracing the full density operator $\hat{\rho}(t)$ of the composite system over the field degrees of freedom:
\begin{equation}
\hat{\rho}_d(t)=\mathrm{Tr}_R\left[\hat{\rho}(t)\right] =\mathrm{Tr}_R\!\left[|\Psi(t)\rangle\langle\Psi(t)|\right].
\label{eq:reduced_rho}
\end{equation}
Here the subscripts ``$d$'' and ``$R$'' denote the detector and the field, respectively.

Combining Eq.~(\ref{eq:reduced_rho}) with Eqs.~(\ref{eq:interaction_H}), (\ref{eq:rho_t}) and (\ref{eq:H_I}), we arrive at
\begin{equation}
\begin{aligned}
|\Psi(t)\rangle &= e^{\alpha(t)}\Big[ c_1 |g\rangle\otimes e^{i\lambda \int_{t_0}^{t}\hat{\bar{\phi}}(t')\, dt'}|0\rangle \\
&\quad+c_2 |e\rangle\otimes e^{-i\lambda \int_{t_0}^{t}\hat{\bar{\phi}}(t')\, dt'}|0\rangle \Big]\;,
\end{aligned}
\label{eq:Psi_t_explicit}
\end{equation}
where 
\begin{equation}
\alpha(t)\equiv-\frac{1}{2}\int_{t_0}^{t} dt_1 \int_{t_0}^{t}dt_2\,[\hat{H}_I(t_2), \hat{H}_I(t_1)]\,\theta(t_2-t_1)
\label{eq:alpha_decoherence}
\end{equation}
with $\theta(x)$ denoting the Heaviside step function. It is easy to check that $\alpha(t)$ is purely imaginary; it therefore only yields a phase factor in the wave function and has no influence on the detector's dynamics.

Substituting Eq.~(\ref{eq:Psi_t_explicit}) into Eq.~(\ref{eq:reduced_rho}) and evaluating the matrix elements of $\hat{\rho}_d(t)$, we get
\begin{equation}
\begin{aligned}
(\rho_d)_{00}&=\langle g|\hat{\rho}_d(t)|g\rangle=|c_1|^2, \\
(\rho_d)_{11}&=\langle e|\hat{\rho}_d(t)|e\rangle=|c_2|^2, \\
(\rho_d)_{01}&=\langle g|\hat{\rho}_d(t)|e\rangle=c_1 c_2^* \, e^{-\gamma(\Delta t)},\\
(\rho_d)_{10}&=\langle e|\hat{\rho}_d(t)|g\rangle =c^*_1 c_2\, e^{-\gamma(\Delta t)},
\end{aligned}
\label{eq:rho_d_matrix}
\end{equation}
where $\Delta t\equiv t-t_0$, and $\gamma(\Delta t)$ is the decoherence function defined as
\begin{equation}
\gamma(\Delta t)\equiv2\lambda^2 \int_{t_0}^{t}dt_1\int_{t_0}^{t}dt_2\;G(t_1,t_2)
\label{eq:gamma_def}
\end{equation}
with
\begin{eqnarray}
G(t_1,t_2)&\equiv&\langle0|\hat{\bar{\phi}}(t_1)\hat{\bar{\phi}}(t_2)|0\rangle\nonumber\\
&=&\int d^3\mathbf{k}\,|g_{\mathbf{k}}|^2 e^{-i\omega_{\mathbf{k}}(t_1-t_2)}
\label{eq:two-point-correlation}
\end{eqnarray}
being the two-point correlation function of the averaged field operator $\hat{\bar{\phi}}(t)$.

Some comments are now in order. First, the diagonal elements of the detector's density matrix remain constant, indicating that the perturbation from vacuum fluctuations produces no net population transfer. This results from the fact that the detector's Hamiltonian $\hat{H}_d=\frac{\omega_0}{2}\,\hat{\sigma}_z$ commutes with the total Hamiltonian of the composite detector-field system, i.e., $[\hat{H}_d,\hat{H}_0+\hat{H}_1]=0$. Second, the off-diagonal elements decay exponentially as $e^{-\gamma(\Delta t)}$, where the decoherence function $\gamma(\Delta t)$, as can be deduced from Eq.~(\ref{eq:gamma_def}), is purely real. This implies that the off-diagonal elements do not acquire any additional dynamical phase during the interaction, and hence the detector undergoes pure decoherence without relative energy-level shifts. This property contrasts sharply with the $\sigma_x$-coupling, where both decoherence and energy shifts coexist~\cite{Breuer02}. Third, the decoherence function $\gamma(\Delta t)$ depends critically on the detector-field coupling strength, the detector profile, and the mode structure of the field (see Eqs.~(\ref{eq:g-function}), (\ref{eq:averaged_field}) and (\ref{eq:gamma_def})). Since the mode structure is highly sensitive to boundary conditions, the decoherence dynamics is expected to behave distinctly under different boundary geometries.


\section{Decoherence dynamics of a detector}

In this section, we investigate the decoherence dynamics of a static detector in three scenarios, i.e., free space, a perfectly reflecting planar boundary, and a cavity formed by two parallel reflecting planar boundaries. 

\subsection{Decoherence dynamics in a free space}
According to Eq.~(\ref{eq:gamma_def}), obtaining the decoherence function in a specific physical setup requires $G(t_1,t_2)$, the two-point correlation function of the averaged field operator. To this end, we first introduce the mode function, in terms of which the averaged field is defined (see Eqs.~(\ref{eq:g-function}) and (\ref{eq:averaged_field})).

In free space, the mode function reads
\begin{equation}
u_{\mathbf{k}}(\mathbf{x})= \frac{e^{i\mathbf{k}\cdot\mathbf{x}}}{\sqrt{(2\pi)^32\omega_{\mathbf{k}}}}\;.
\label{eq: mode_no_boundary}
\end{equation}
Substituting this expression along with Eq.~(\ref{eq:form_factor}) into Eq.~(\ref{eq:g-function}) and doing the integrations, we obtain
\begin{equation}
g_{\mathbf{k}}
=\frac{e^{-\omega_{\mathbf{k}}l}}{4\pi\sqrt{\pi\omega_{\mathbf{k}}}},
\label{eq:gk_explicit}
\end{equation}
which exhibits an exponential decay with respect to the mode frequency  $\omega_{\mathbf{k}}$.
This character further carries onto the averaged field operator $\hat{\bar{\phi}}(t)$, the use of which in Eq.~(\ref{eq:two-point-correlation}) then yields:
\begin{equation}
G_0(t_1,t_2)= -\frac{1}{4\pi^2}\frac{1}{\left(t_1-t_2-2il\right)^2}\;.
\label{eq:two_point_no_boundary}
\end{equation}
It shows that the detector's profile, characterized by the parameter $l$, effectively regularizes the two-point correlation function, which diverges in the limit $l\rightarrow0$.

Inserting the above $G_0(t_1,t_2)$ into Eq.~(\ref{eq:gamma_def}) and performing the double integral, we get
\begin{align}
\gamma_0(\Delta t)
&= \frac{\lambda^2}{2\pi^2}\ln\left(1+\frac{\Delta t^2}{4l^2}\right),
\label{eq:gamma_0}
\end{align}
which is the decoherence function of a detector in free space. Since $\Delta t>0$, it follows that $\gamma_0(\Delta t)$ is always positive; moreover, it is a monotonically increasing function of $\Delta t$. Accordingly, the off-diagonal matrix elements, $(\rho_d)_{01}=(\rho_d)^{\ast}_{10}\propto e^{-\gamma_0(\Delta t)}$, decay monotonically with $\Delta t$. When $\Delta t\rightarrow\infty$, they tend to zero, indicating that the detector evolves from a superposition state with quantum coherence
to a mixture with classical probability distribution,
as a result of the perturbation of vacuum fluctuations. To quantitatively characterize this process, we introduce the decoherence time $\Delta t_c$, defined as the characteristic time required for the off-diagonal elements to decay to $e^{-1}$ of their initial value. 
We find that $\Delta t_c= 2l\sqrt{e^{2\pi^2/\lambda^2}-1}$. Therefore, a larger detector size $l$ or a weaker coupling $\lambda$ leads to slower decoherence.

\subsection{Decoherence dynamics near a perfectly reflecting boundary}

\begin{figure}[H]
\centering
\includegraphics[width=0.8\columnwidth]{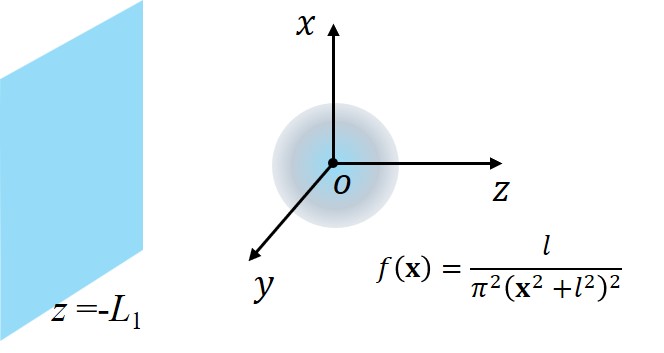}
\caption{Schematic of a detector near a perfectly reflecting planar boundary.}
\label{fig:oneplate}
\end{figure}

As shown in Fig.~\ref{fig:oneplate}, we consider a static detector placed near a perfectly reflecting infinite planar boundary. The coordinate origin is chosen at the detector center, such that the boundary lies at $z=-L_1$. We assume that the detector-boundary distance $L_1$ is much larger than the detector's characteristic size $l$, and that the boundary-induced modification of the detector profile can be neglected. 

Imposing the Dirichlet boundary condition at the boundary $z=-L_1$, we express the mode function as:
\begin{eqnarray}
u_{\mathbf{k}}(\mathbf{x})=
\frac{2e^{i\mathbf{k}_\parallel\cdot\mathbf{x}_\parallel}\sin\left[k_z(z+L_1)\right]}{\sqrt{(2\pi)^32\omega_{\mathbf{k}}}}\theta(k_z)\theta(z+L_1)\;,
\label{eq:field-mode_plate}
\end{eqnarray}
where $\mathbf{x}=(\mathbf{x}_\parallel, x_z)$, $\mathbf{k}=(\mathbf{k}_\parallel, k_z)$, and the subscript ``$\parallel$" denotes components of vectors transverse to the $z$-axis. 

Substituting this mode function along with the profile function, Eq.~(\ref{eq:form_factor}), into Eq.~(\ref{eq:g-function}) and performing the integration, we obtain
\begin{equation}
g_{\mathbf{k}}=\frac{\sin(k_z L_1)e^{-\omega_{\mathbf{k}}l}}{2\pi\sqrt{\pi\omega_{\mathbf{k}}}}\cdot\theta(k_z),
\label{eq:gk_plate_integral}
\end{equation}
which further determines the averaged field operator $\hat{\bar{\phi}}(t)$ through Eq.~(\ref{eq:averaged_field}). Then the two-point correlation function for the averaged field, 
follows as
\begin{eqnarray}
G_1(t_1,t_2)
&=&-\frac{1}{4\pi^2}\frac{1}{(t_1-t_2-2il)^2}\nonumber\\&&
+\frac{1}{4\pi^2}\frac{1}{(t_1-t_2-2il)^2-4L_1^2}\;.
\label{eq:two_point_plate}
\end{eqnarray}
It consists of two terms: the first term is independent of the detector-boundary separation $L_1$ and identical to its free-space counterpart (see Eq.~(\ref{eq:two_point_no_boundary})), whereas the second term depends strongly on $L_1$ and thus arises from the presence of the boundary. Specifically, the second term describes the correlation between the averaged field operator and its image counterpart. 

Combining Eq.~(\ref{eq:two_point_plate}) with Eq.~(\ref{eq:gamma_def}) and simplifying, we arrive at the following decoherence function: 
\begin{equation}
\gamma_1(\Delta t,L_1)
=\gamma_0(\Delta t)+\delta\gamma_1(\Delta t,L_1),
\label{eq:gamma_plate}
\end{equation}
where $\gamma_0(\Delta t)$ is the decoherence function in free space (see Eq.~(\ref{eq:gamma_0})), and
\begin{eqnarray}
&&\delta\gamma_1(\Delta t,L_1)\nonumber \\
&=&\frac{\lambda^2}{2\pi^2}
\Bigg\{-\frac{\Delta t}{4L}\ln\frac{\left(\Delta t+2L_1\right)^2+4l^2}{\left(\Delta t-2L_1\right)^2+4l^2}\nonumber\\
&&-\ln\frac{\sqrt{\left(\left(\Delta t+2L_1\right)^2+4l^2\right)\left(\left(\Delta t-2L_1\right)^2+4l^2\right)}}{4(L_1^2+l^2)}\nonumber \\
&&
+\frac{l}{L}\Bigg[2\arctan\left(\frac{L_1}{l}\right)-\arctan\left(\frac{\Delta t+2L_1}{2l}\right)\nonumber\\&&
-\arctan\left(\frac{2L_1-\Delta t}{2l}\right)\Bigg]\Bigg\}\;.
\label{eq:gamma_plate_explicit}
\end{eqnarray}
This result shows that, in contrast to the free-space case, the decoherence function near a perfectly reflecting boundary acquires an extra boundary correction term $\delta\gamma_1(\Delta t, L_1)$. 

Taking the limit $\Delta t\rightarrow0$, it is straightforward to verify that $\gamma_1(\Delta t,L_1)\rightarrow0$, as does the decoherence function in the free-space case. This is consistent with the initial condition  $\rho_{01}(t_0)=\rho^{\star}_{10}(t_0)=c_1c_2^{\star}$. 

When $\Delta t\rightarrow\infty$, 
\begin{eqnarray}
\gamma_1^f(L_1)\equiv\lim_{\Delta t\rightarrow\infty}\gamma_1(\Delta t,L_1)\quad\quad\quad\quad\quad\quad\quad\quad\quad\quad\quad\quad\nonumber\\\;\;
\approx\frac{\lambda^2}{2\pi^2}\left\{\ln\left[1+\left(\frac{L_1}{l}\right)^2\right]+2\frac{l}{L_1}\arctan\left({\frac{L_1}{l}}\right)-2\right\}.\;\;
\label{gamma1f}
\end{eqnarray}
Since $L_1>l$, $\gamma_1^f(L_1)$ remains positive, so the decoherence function approaches a positive constant in the large-time limit. This contrasts sharply with the free-space case, where the decoherence function grows without bound as $\Delta t\rightarrow\infty$. Consequently, the off-diagonal elements of the density matrix at late times behave as $e^{-\gamma^f_{1}(L_1)}$, 
indicating that the coherence of a detector near the boundary is partially preserved during the evolution, unlike the free-space case where vacuum fluctuations completely destroy coherence.

\begin{figure*}[!htbp]
\centering
\begin{subfigure}[b]{0.49\textwidth}
\centering
\includegraphics[width=\linewidth]{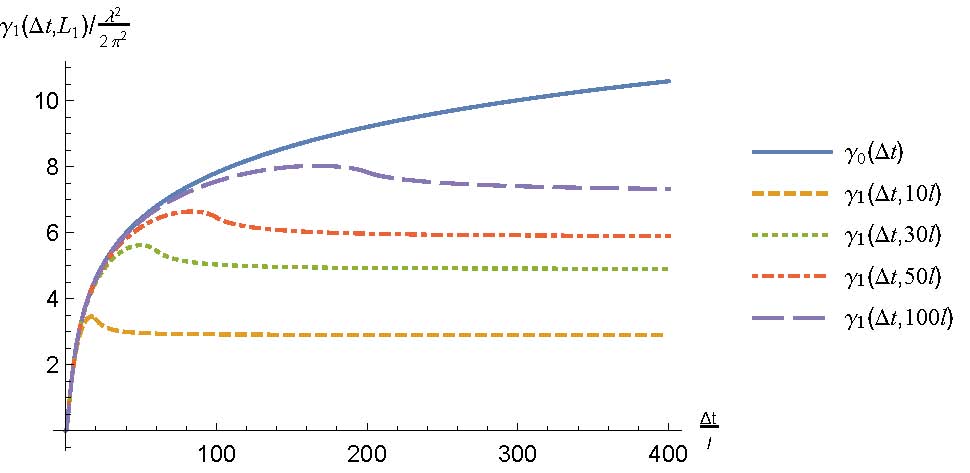}
\caption{Decoherence function near a boundary vs its counterpart in free space.}
\label{Fig1b}
\end{subfigure}
\hfill
\begin{subfigure}[b]{0.49\textwidth}
\centering
\includegraphics[width=\linewidth]{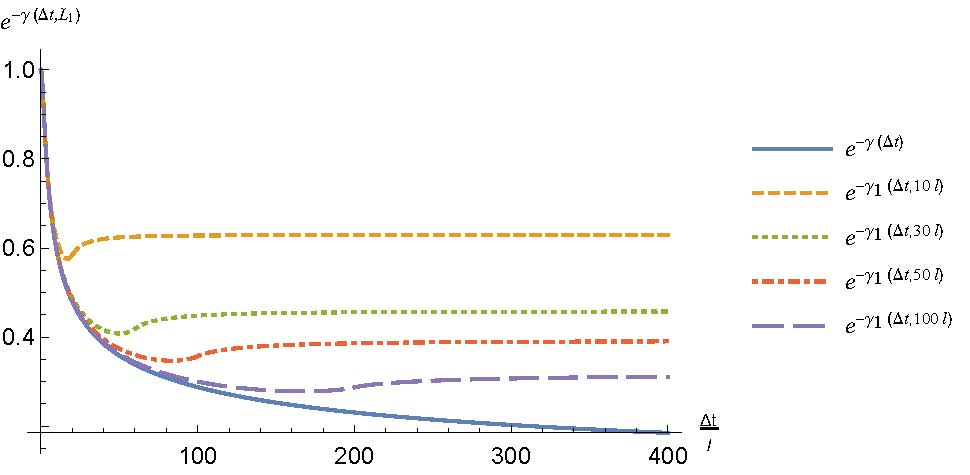}
\caption{Decay of off-diagonal components of the detector's density matrix. The coupling constant is set to be $\lambda=1.0$.}
\label{Fig2b}
\end{subfigure}
\captionsetup{justification=justified, singlelinecheck=false}
\caption{Decoherence dynamics of a detector near a perfectly reflecting boundary. The vertical axis in subfigure~(\ref{Fig1b}) denotes the decoherence function in units of $\frac{\lambda^2}{2\pi}$. In both subfigures, the horizontal axes represent the time interval $\Delta t$ in units of $l$; the solid blue lines correspond to the free-space case, while the dashed yellow, dotted green, dash-dotted red, and  thick-dashed light-blue lines correspond to detector-boundary distances $L_1=10l, 30l, 50l$, and $100l$, respectively.}
\label{Fig1b2b}
\end{figure*}

For the full decoherence dynamics, the analytical expressions given in Eqs.~(\ref{eq:gamma_plate}) and (\ref{eq:gamma_plate_explicit}) become too complicated to provide immediate physical insights. We therefore resort to numerical analysis. Figure~\ref{Fig1b2b} shows the decoherence dynamics of a detector at various distances from the boundary: the dashed yellow,   dotted green, dash-dotted red, and thick-dashed  light-blue lines correspond to distances $L_1=10l, 30l, 50l, 100l$, while the heavy solid blue line represents the free-space case. Several prominent features of the decoherence dynamics emerge from this comparison.

First, as shown in subfigure (\ref{Fig1b}), the decoherence function deviates significantly from that in the free-space case when the detector-boundary separation is relatively small (for instance, see the dashed yellow line), and the larger the detector-boundary separation, the smaller the discrepancy between the decoherence functions in the two cases. This behavior can also be verified analytically. In the limit of a very large detector-boundary separation, 
\begin{equation}
\gamma_1(\Delta t,L_1)\Big|_{L_1\rightarrow\infty}	\approx\frac{\lambda^2}{2\pi^2}\ln\left(1+\frac{\Delta t^2}{4l^2}\right)-\frac{\lambda^2\Delta t^2}{8\pi^2L_1^2}\;,
\label{eq:gamma_plate_largeL}
\end{equation}
where the second term (the boundary correction) is negligibly small compared with the first term, so that the free-space result is recovered. This is also expected on physical grounds, since the boundary effect vanishes as the detector-boundary separation tends to infinity. 

Second, unlike the free-space case, where the decoherence function 
increases monotonically with time, the decoherence function near the boundary first increases with time, reaches a maximum, and then decays very slowly to a constant value. This final constant value agrees with the earlier analytical result (see Eq.~(\ref{gamma1f})). 
As can be observed from  Fig.~\ref{Fig1b2b}, the decoherence time, i.e., the time required for  $e^{-\gamma_1(\Delta t,L_1)}$ to reach its stable asymptotic value, becomes shorter as the detector-boundary separation decreases.

Third, the factor $e^{-\gamma_1(\Delta t,L_1)}$ characterizing the dynamical evolution is always larger than its free-space counterpart, indicating that the boundary suppresses decoherence. 

\begin{figure}[H]
\centering
\includegraphics[width=1.0\columnwidth]{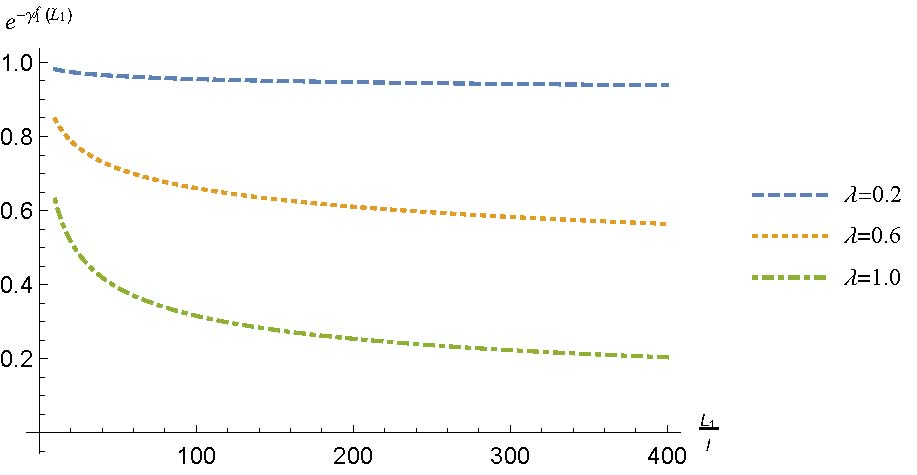}
\caption{Asymptotic coherence for a detector located at various distances from a perfectly reflecting boundary. The dashed blue, dotted yellow, and dash-dotted green lines represent $\lambda=0.2, 0.6$, and $1.0$, respectively.}
\label{fig:Fig3b}
\end{figure}

We adopt the factor $e^{-\gamma_1^f(L_1)}$, the ratio of the final to the initial value of the off-diagonal elements of the detector's density matrix, as a quantitative measure of coherence preservation after long-time evolution. Equation (\ref{gamma1f}) shows that this factor depends crucially on the coupling constant $\lambda$ and the detector-boundary separation $L_1$. As shown by the dashed blue line in Fig.~\ref{fig:Fig3b}, which corresponds to a relatively small coupling constant, $\lambda=0.2$, $e^{-\gamma_1^f(L_1)}$ is slightly smaller than unity and decreases marginally with increasing $L_1$, showing that decoherence of the detector can be effectively suppressed by the presence of the boundary. By contrast, for larger values of $\lambda$, corresponding to stronger detector-field coupling (see the yellow and green lines), $e^{-\gamma_1^f(L_1)}$ decreases more obviously with increasing $L_1$. Moreover, for a fixed detector-boundary separation, a smaller coupling constant yields larger values of $e^{-\gamma_1^f(L_1)}$, i.e., weaker detector-field coupling leads to less decoherence. In summary, weaker detector-field coupling and shorter detector-boundary separation are favorable for preserving the detector's coherence. 

\subsection{Decoherence dynamics in a cavity}
\begin{figure}[]
\centering
\includegraphics[width=0.7\columnwidth]{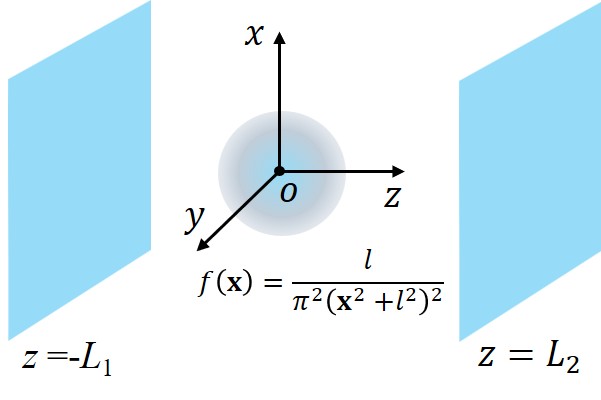}
\caption{Schematic of a detector placed inside a cavity formed by two parallel perfectly reflecting  boundaries. The detector is located at the origin, and the boundaries at $z=-L_1$ and $z=L_2$, with $L=L_1+L_2$ the cavity width.}
\label{fig:twoplates}
\end{figure}

We now explore the decoherence dynamics of a detector placed inside a cavity, formed by two parallel perfectly reflecting boundaries. As shown in Fig.~\ref{fig:twoplates}, the detector is placed at the origin and the two boundaries are located at $z=-L_1$ and $z=L_2$, respectively, where $L_1,L_2>0$ and $L=L_1+L_2$.

We assume that the cavity width and the two detector-boundary separations are much larger than the size of the detector, i.e., $L, L_1, L_2\gg l$.    
At the two boundaries, the field satisfies the Dirichlet boundary condition,  $\phi(t,\mathbf{x})|_{z=-L_1}=\phi(t,\mathbf{x})|_{z=L_2}=0$. The field operator can then be expanded in terms of a complete set of field modes:
\begin{eqnarray}
\phi(t,\mathbf{x})=\sum_{n=1}^{\infty} \int_{\mathbb{R}^2} d^2\mathbf{k}_\parallel\Big[e^{-i\omega_n t}u_{\mathbf{k}_\parallel,k_{zn}}(\mathbf{x})a_{\mathbf{k}_\parallel,k_{zn}}
+\text{H.c.}\Big],\nonumber\\
\label{eq:field_cavity}
\end{eqnarray}
where   
\begin{equation}
u_{\mathbf{k}_\parallel,k_{zn}}(\mathbf{x})=\frac{e^{i\mathbf{k}_\parallel\cdot\mathbf{x}_\parallel}\sin\left(k_{zn}\left(z+L_1\right)\right)}{2\pi\sqrt{\omega_nL}}\theta\left(\left(z+L_1\right)\left(L_2-z\right)\right)
\end{equation}
with $\omega_n=\sqrt{k_\parallel^2+k_{zn}^2}$, $k_\parallel=|\mathbf{k}_{\parallel}|$,  $k_{zn}=\frac{\pi n}{L}$, and $n$ an integer ranging from $0$ to $\infty$. 

Substituting this mode function and the detector's profile function (Eq.~(\ref{eq:form_factor}) into Eq.~(\ref{eq:g-function}) yields
\begin{eqnarray}
g_{\mathbf{k}_\parallel,k_{zn}}&=&\frac{k_{\parallel}l}{2\pi^2\sqrt{\omega_nL}}\int^{L_2}_{-L_1}dz\frac{\sin\left(k_{zn}\left(z+L_1\right)\right)}{\sqrt{z^2+l^2}}\nonumber\\&&\times K_1\left(k_{\parallel}\sqrt{z^2+l^2}\right)\;,
\label{eq:gk-cavity}
\end{eqnarray}
where $K_{\nu}(x)$ is the modified Bessel function of the second kind. 

Inserting $g_{\mathbf{k}_\parallel,k_{zn}}$ into Eq.~(\ref{eq:two-point-correlation}), we obtain the averaged field operator $\hat{\bar{\phi}}(t)$, whose two-point correlation function is then given by 
\begin{eqnarray}
&&G_2(t_1,t_2)=\frac{l^2}{4\pi^4L}\sum_{n=1}^{\infty}\int_{\mathbb{R}^2} d^2\mathbf{k}_{\parallel}\frac{k^2_{\parallel}}{\omega_n}e^{-i\omega_n(t_1-t_2)}\quad\quad\nonumber\\&&
\times\left|\int^{L_2}_{-L_1}dz\frac{\sin\left(k_{zn}\left(z+L_1\right)\right)}{\sqrt{z^2+l^2}}\cdot K_1\left({k}_{\parallel}\sqrt{z^2+l^2}\right)\right|^2.\quad\quad
\end{eqnarray}

Next, we substitute this function into Eq.~(\ref{eq:gamma_def}), perform the $t_1$- and $t_2$-integrals, and arrive at
\begin{eqnarray}
\gamma_2(\Delta t,L_1,L)=\frac{\lambda^2l^2}{\pi^4L}\sum_{n=-\infty}^{\infty}\int_{\mathbb{R}^2} d^2\mathbf{k}_{\parallel}\frac{k^2_{\parallel}}{\omega_n}\sin^2\left(\frac{\omega_n}{2}\Delta t\right)\nonumber\\
\times\left|\int^{L_2}_{-L_1}dz\frac{\sin\left(k_{zn}\left(z+L_1\right)\right)}{\sqrt{z^2+l^2}}\cdot K_1\left({k}_{\parallel}\sqrt{z^2+l^2}\right)\right|^2.\;\;
\end{eqnarray}

\begin{figure}[]
\centering
\includegraphics[width=1.0\columnwidth]{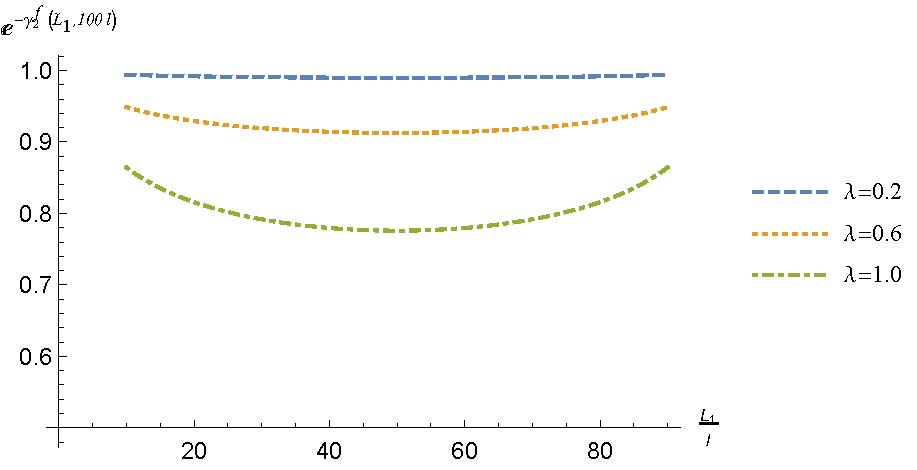}
\caption{Asymptotic coherence for a detector in a cavity, located at a distance $L_1\in[10l,90l]$ from one of the boundaries. The cavity width is set to be $L=100l$.}
\label{fig:Figcavity0}
\end{figure}

Assuming that both $L_1$ and $L_2$ are at least an order of magnitude larger than the detector's size parameter $l$, we can further simplify the above decoherence function into
\begin{eqnarray}
\gamma_2(\Delta t,L_1,L)&=&\frac{4\lambda^2}{\pi L}\sum_{m=1}^{\infty}\sin^2\left(\frac{m\pi L_1}{L}\right)\int^{\infty}_{\frac{\pi m}{L}}d\omega\;\frac{e^{-2\omega l}}{\omega^{2}}\nonumber\\&&\times
\sin^2\left(\frac{\omega}{2}\Delta t\right)\;.
\end{eqnarray}
This is the general expression for the decoherence function of a detector located at distances $L_1$ and $L_2=L-L_1$ from the two boundaries. As in the single-boundary case, this function vanishes when $\Delta t=0$ and approaches a constant as $\Delta t\rightarrow\infty$:
\begin{eqnarray}
\gamma^f_2(L_1,L)&\equiv&\lim_{\Delta t\rightarrow\infty}\gamma_2(\Delta t,L_1,L)\nonumber\\&
\approx&\frac{2\lambda^2}{\pi L}\sum_{m=1}^{\infty}\sin^2\left(\frac{m\pi L_1}{L}\right)\int^{\infty}_{\frac{\pi m}{L}}d\omega\;\frac{e^{-2\omega l}}{\omega^2}.\quad\;\;\;
\end{eqnarray}
This non-vanishing result suggests that, just as in the single-boundary case, the coherence of a detector can also be partially preserved in the cavity. We can see from this expression that $\gamma_2^f(L_1,L)$, and thus the corresponding decoherence factor $e^{-\gamma_2^f(L_1,L)}$, depends crucially on the detector-field coupling constant $\lambda$, the cavity width $L$, and $L_1$ characterizing the relative position of the detector with respect to the cavity. As shown in Fig.~\ref{fig:Figcavity0}, for fixed values of the cavity width $L$ and detector-boundary separation $L_1$, stronger coupling leads to a smaller decoherence factor. Moreover, this factor displays a symmetric behavior with respect to the detector-boundary separation $L_1$: the further the detector is from the cavity center, the larger the decoherence factor. 

Keeping $L_1$ finite and letting $L_2$, and hence $L$, tend to infinity, the above decoherence function reduces to  
\begin{eqnarray}
\gamma_2(\Delta t,L_1)&\equiv&\lim_{L_2\rightarrow\infty}\gamma_2(\Delta t,L_1,L)\nonumber\\&
\approx&\frac{2\lambda^2}{\pi^2}\int^{\infty}_0d\omega\;\frac{e^{-2\omega l}}{\omega}\sin^2\left(\frac{\omega}{2}\Delta t\right)\nonumber\\&-&\frac{\lambda^2}{\pi^2L_1}\int^{\infty}_0d\omega\frac{e^{-2\omega l}}{\omega^2}\sin(2\omega L_1)\sin^2\left(\frac{\omega}{2}\Delta t\right).\nonumber\\
\end{eqnarray}
Carrying out the integrals further, we find that this decoherence function is exactly that of a detector near a single boundary, i.e., $\gamma_2(\Delta t,L_1)=\gamma_1(\Delta t,L_1)$.
Therefore, in the limit $L_2\rightarrow\infty$ with finite $L_1$, the decoherence function of a detector in the cavity, $\gamma_2(\Delta t,L_1)$, reduces to $\gamma_1(\Delta t,L_1)$, the decoherence function of a detector at a distance $L_1$ from a perfectly reflecting boundary. Similarly, the single-boundary result is recovered when $L_1\rightarrow\infty$ with $L_2$ finite. 

\begin{figure*}[htbp]
\centering
\begin{subfigure}[b]{0.49\textwidth}
\centering
\includegraphics[width=\linewidth]{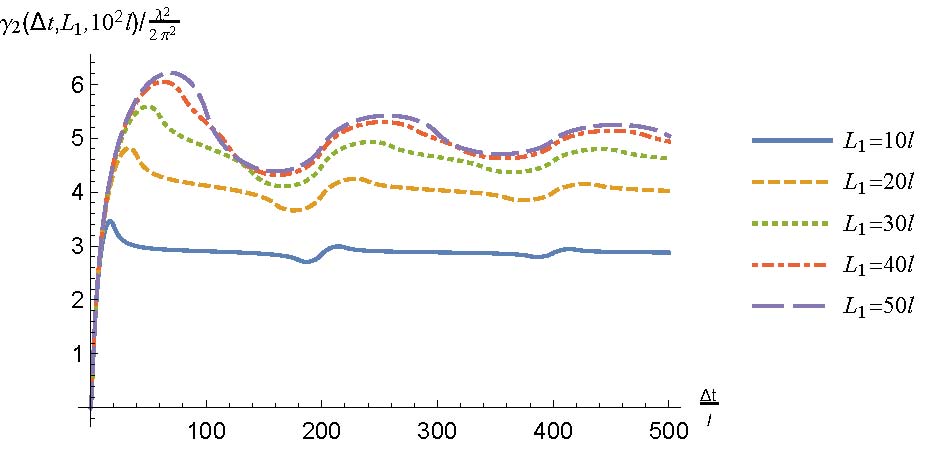}
\caption{Decoherence function of a detector in a cavity.}
\label{Figcavity1}
\end{subfigure}
\hfill
\begin{subfigure}[b]{0.49\textwidth}
\centering
\includegraphics[width=\linewidth]{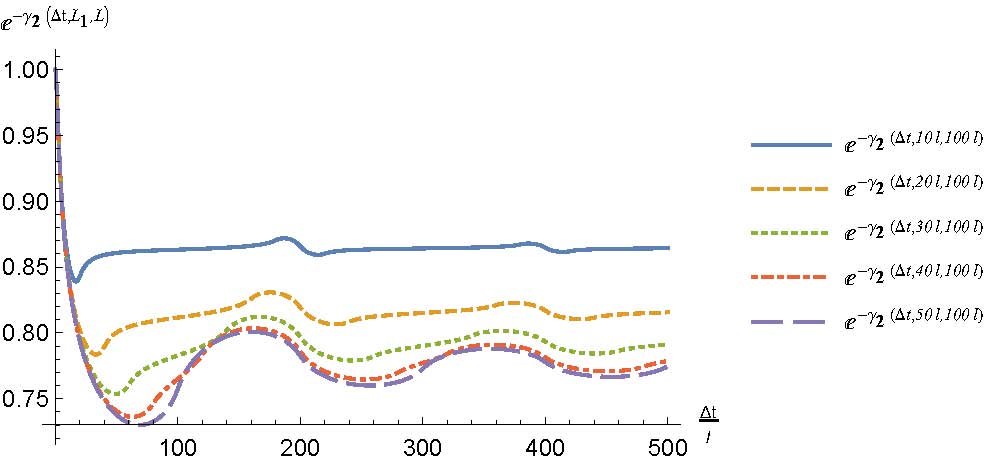}
\caption{Decay of off-diagonal elements of the detector's density matrix in a cavity. The coupling constant is $\lambda=1.0$.}
\label{Figcavity2.pdf}
\end{subfigure}
\caption{Decoherence dynamics of a detector in a cavity of width $L=100l$. The left and right subfigures display the decoherence function $\gamma_2(\Delta t,L_1,L)$ in units of $\frac{\lambda^2}{2\pi}$ and the decoherence factor $e^{-\gamma_2(\Delta t,L_1,L)}$, respectively. The horizontal axes in both subfigures represent the time interval in units of $l$. In both subfigures, the solid blue, dashed yellow, dotted green, dash-dotted red, and wide-dashed light-blue lines correspond to the detector-boundary distances $L_1=10l, 20l, 30l, 40l$, and $50l$, respectively.}
\label{Figcavity}
\end{figure*}

We next turn to numerical results to illustrate the effects of the cavity on the detector's decoherence dynamics. Figure~\ref{Figcavity} displays the dynamical evolution of the detector's coherence inside a cavity of width $L=100l$, for various detector positions characterized by the distance $L_1$ to the nearest boundary. In sharp contrast to both the free-space and single-boundary cases, the decoherence dynamics clearly exhibits oscillatory behavior, with the oscillation amplitude decaying over time and increasing with $L_1$.  

\begin{figure}[H]
\centering
\includegraphics[width=1.0\columnwidth]{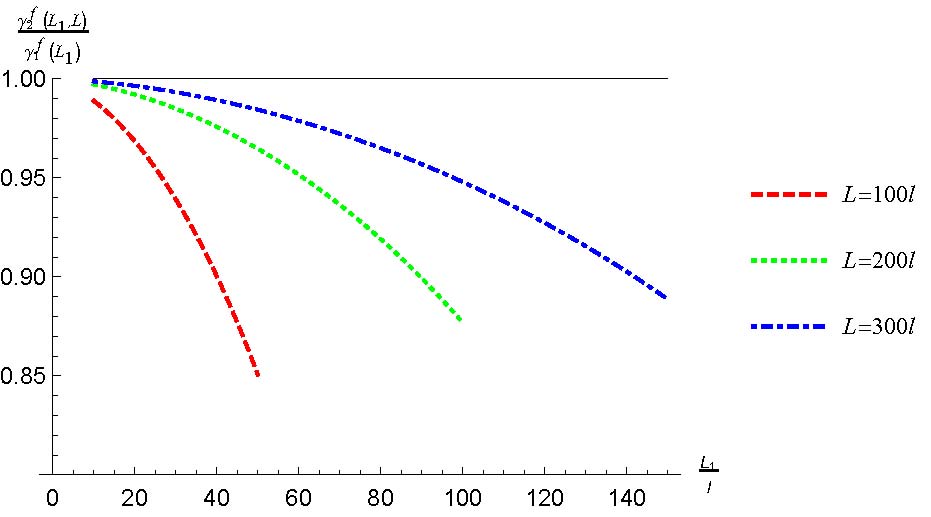}
\caption{Dependence of the ratio between the decoherence functions in the cavity and single-boundary cases, $\frac{\gamma_2^f(L_1,L)}{\gamma_1^f(L_1)}$, on the nearest detector-boundary separation $L_1$, for cavity width $L=100l, 200l$, and $300l$. The dashed red, dotted green, and dash-dotted blue lines correspond to these three widths, respectively.}
\label{fig:Figratio0}
\end{figure}

Lastly, we compare the asymptotic coherence for a detector at the same distance $L_1$ from the nearest boundary in the cavity and single-boundary cases, characterized by  $e^{-\gamma^f_2(L_1,L)}$ and $e^{-\gamma^f_1(L_1)}$, respectively. As shown in Fig.~\ref{fig:Figratio0}, for a detector equidistant from the nearest boundary in both geometries, the ratio of the decoherence functions, $\frac{\gamma_2^f(L_1,L)}{\gamma_1^f(L_1)}$, is always less than unity. 
Consequently, $e^{-\gamma^f_2(L_1,L)}>e^{-\gamma^f_1(L_1)}$, i.e., more coherence is preserved in the cavity case than in the single-boundary case.

\section{Summary}

In this paper, we investigate boundary effects on the decoherence dynamics of a two-level system (detector) coupled to a massless scalar field in vacuum via the $\hat{\sigma}_z$-coupling scheme. By rigorously evaluating the matrix elements of the reduced density operator, we find that only the off-diagonal elements of the detector's density matrix evolve in time, whereas the diagonal elements remain unchanged. This indicates that the $\hat{\sigma}_z$-coupling induces no net population transfer, but purely governs the evolution of the detector's coherence. Specifically, the off-diagonal elements decay as $e^{-\gamma(\Delta t)}$, with the decoherence function $\gamma(\Delta t)$ defined in terms of the two-point correlation function of the averaged field operator (see Eq.~(\ref{eq:gamma_def})). On this basis, we explicitly examine the decoherence dynamics of a detector in three scenarios: free space, a perfectly reflecting boundary, and a cavity formed by two parallel perfectly reflecting boundaries. 

In the free-space case, the decoherence function varies monotonically with $\Delta t$, and consequently the detector's coherence decays monotonically as well. The decoherence rate is highly sensitive to the coupling constant $\lambda$ and the detector's size parameter $l$: weaker coupling or a larger detector yields a slower decoherence rate. For a detector near a perfectly reflecting boundary, the decoherence function consists of two terms: a position-independent term identical to its free-space counterpart, and a strongly position-dependent term originating from the presence of the boundary. Consequently, the decoherence function displays non-monotonic behavior. Accordingly, the off-diagonal elements of the detector's density matrix first decrease in time to a minimum and then rise slightly to a stable value, indicating that the initial coherence can be partially preserved owing to the presence of the boundary. The closer the detector is to the boundary, the more coherence can be preserved. For a detector inside a cavity, the decoherence function, and hence the decoherence dynamics, exhibits oscillatory behavior, with the oscillation amplitude decaying in time. As in the single-boundary case, the initial coherence of the detector can be partially preserved at long times: the smaller the distance from the detector to the nearest boundary, the more coherence can be preserved. However, when the detector is located at the cavity center, the boundary-induced suppression of decoherence reaches its minimum. In addition, for a detector at the same distance from the nearest boundary, more coherence is preserved in the cavity case than in the single-boundary case.

These results suggest that the decoherence dynamics can be effectively manipulated via boundary engineering. In the absence of boundaries, vacuum fluctuations inevitably lead to a complete loss of coherence. In contrast, the presence of a single boundary or a cavity reshapes the field modes and modifies the vacuum fluctuation correlations, thereby altering the decoherence function and allowing the initial coherence to be partially preserved. Furthermore, the preserved coherence depends sensitively on the boundary configuration and detector position, with positions closer to the boundary retaining more coherence. These findings indicate that boundary engineering offers a promising and experimentally accessible route to manipulating quantum coherence in quantum systems.

\begin{acknowledgments}
This work was supported in part by the Zhejiang Provincial Natural Science Foundation of China under Grant No. LY24A050001, the Ningbo Science $\&$ Technology Bureau under program grant No. 2024DJRC010009, and the K. C. Wong Magna Fund in Ningbo University.
\end{acknowledgments}

\vspace{6cm}
\bibliography{apssamp}

\end{document}